\documentclass[%
 reprint,
 amsmath,amssymb,
 aps,
]{revtex4-1}

\newcommand{\mc}{\mathcal}
\usepackage{color}

\usepackage{hyperref}
\usepackage{graphicx}
\usepackage{dcolumn}
\usepackage{bm}
\usepackage{color}

\begin{document}

\preprint{APS/123-QED}

\title{Small-World Brain Networks Revisited}

\author{Danielle S. Bassett$^{1,2}$}
\author{Edward T. Bullmore$^{3,4}$}
\affiliation{
 $^1$Department of Bioengineering, University of Pennsylvania, Philadelphia, PA, 19104
}
\affiliation{
$^2$Department of Electrical and Systems Engineering, University of Pennsylvania, Philadelphia, PA, 19104
}
\affiliation{
 $^3$Department of Psychiatry, University of Cambridge, Cambridge UK CB2 0SZ
}
\affiliation{$^4$ImmunoPsychiatry, Immuno-Inflammation Therapeutic Area Unit, GlaxoSmithKline R\&D, Stevenage UK SG1 2NY}

\date{\today}
\begin{abstract}
It is nearly 20 years since the concept of a small-world network was first quantitatively defined, by a combination of high clustering and short path length; and about 10 years since this metric of complex network topology began to be widely applied to analysis of neuroimaging and other neuroscience data as part of the rapid growth of the new field of connectomics. Here we review briefly the foundational concepts of graph theoretical estimation and generation of small-world networks. We take stock of some of the key developments in the field in the past decade and we consider in some detail the implications of recent studies using high-resolution tract-tracing methods to map the anatomical networks of the macaque and the mouse. In doing so, we draw attention to the important methodological distinction between topological analysis of binary or unweighted graphs, which have provided a popular but simple approach to brain network analysis in the past, and the topology of weighted graphs, which retain more biologically relevant information and are more appropriate to the increasingly sophisticated data on brain connectivity emerging from contemporary tract-tracing and other imaging studies. We conclude by highlighting some possible future trends in the further development of weighted small-worldness as part of a deeper and broader understanding of the topology and the functional value of the strong and weak links between areas of mammalian cortex.
\end{abstract}

\maketitle

\section*{Small-worlds, Watts and Strogatz}
\begin{figure*}[h]
\begin{center}
\includegraphics[width=0.7\textwidth]{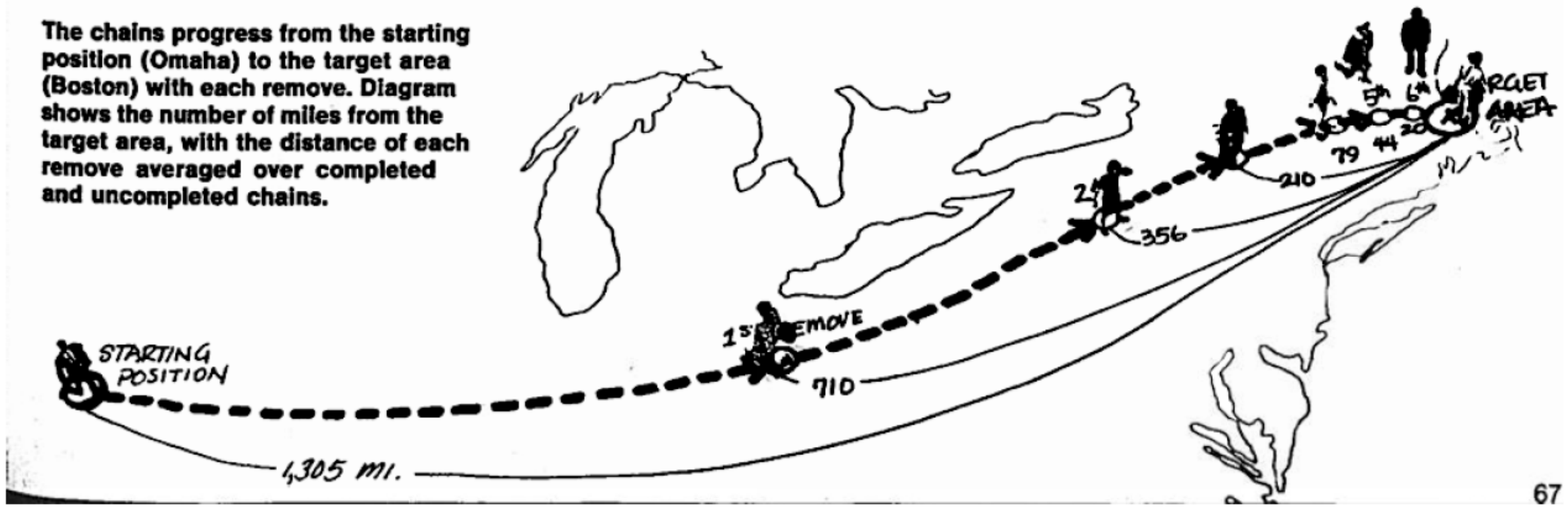}
\caption{ \textbf{An illustration of the shortest path between Omaha and Boston in Milgram's social network experiment.} An image from Stanley Milgram's original experiment, published in {\em Psychology Today} in 1967. Here, the results of multiple experiments are represented as a composite shortest path between the source (a person in Omaha) and the target (a person in Boston). A letter addressed to the target was given to the source, who was asked to send it on (with the same instructions) to the friend or acquaintance that they thought was most likely to know the target, or someone else who might know the target personally. It was found that most letters that eventually reached the correct address in Boston passed through six intermediaries between source and target (denoted 1st remove, 2nd remove, etc), popularising the notion that each of us is separated by no more than ``six degrees of freedom'' from any other individual in a geographically distributed social network.  Reproduced with permission from \cite{Milgram1967}. \label{Milgram}}
\end{center}
\end{figure*}

Small-worldness now seems to be a ubiquitous characteristic of many complex systems; but its first, and still most familiar, appearance was in the form of social networks. We know that as individual agents (nodes) in a social network, we are connected by strong familial and friendship ties (edges) to a relatively few people who are likely also strongly connected to each other, forming a social clique, family or tribe. Yet we also know that we can travel far away from our tribal network, to physically remote cultures and places, and sometimes be surprised there to meet people -- often ``friends-of-friends'' -- who are quite closely connected to our home tribe: ``it's a small world'', we say. This common intuition was experimentally investigated by Milgram, who asked people in the mid-West of the US (Omaha, Nebraska) to forward a letter addressed to an unknown individual in Boston by posting it to the friend or acquaintance in their social network that they thought might know someone else who would know the addressee \cite{Milgram1967} (Fig.~\ref{Milgram}). It was discovered, on average over multiple trials of this procedure, that the letters successfully reaching Boston had been passed through 6 intermediate postings, which was considered much less than expected given the geographical distance between source and target addresses. In the language of graph theory, the characteristic path length of Milgram's social networks was short.

\begin{figure*}[h]
\begin{center}
\includegraphics[width=0.5\textwidth]{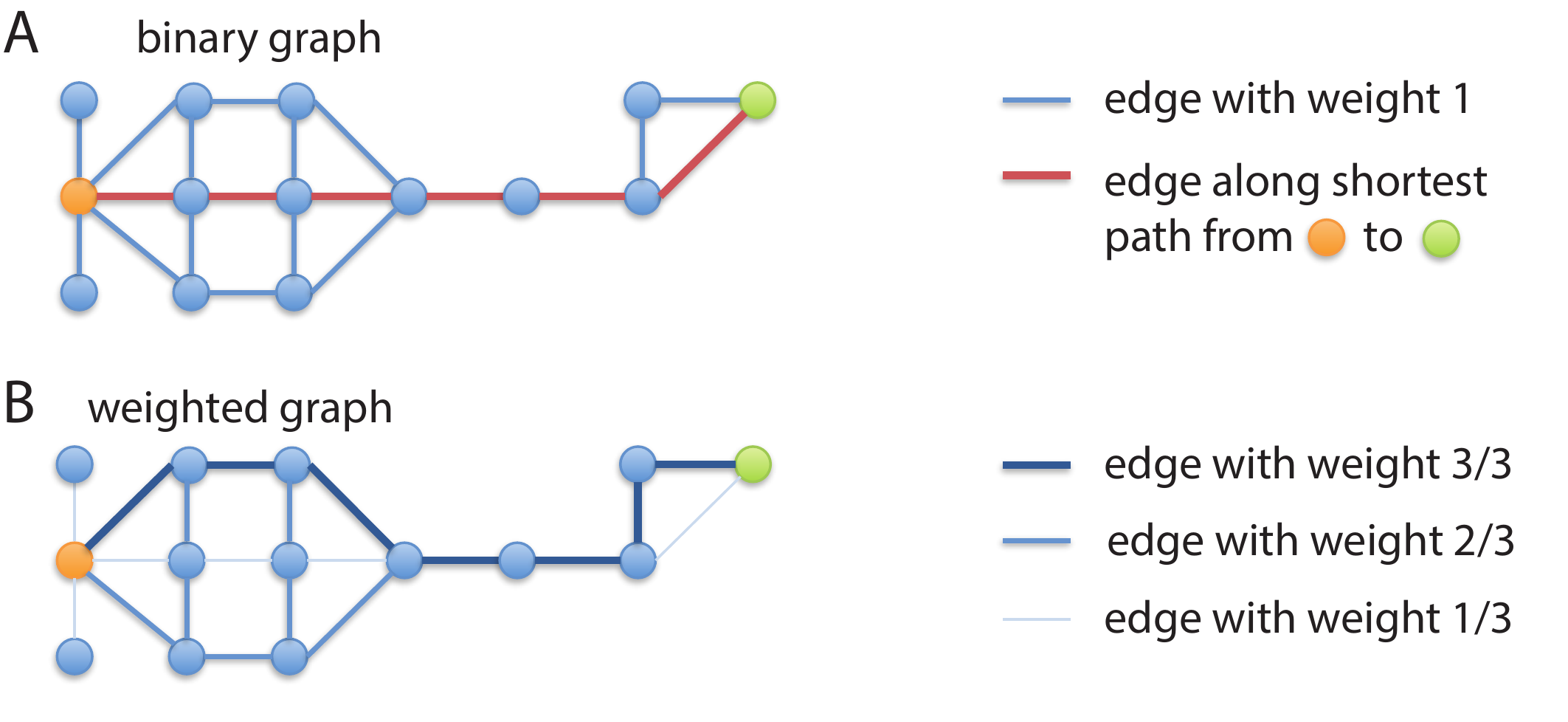}
\caption{ \textbf{Diagrams of clustering and path length in binary and weighted networks.} \emph{(A)} In a binary network, all edges have the same weight, and that is a weight equal to unity. In this example of a binary graph, if one wishes to walk along the shortest path from the orange node to the green node, then one would choose to walk along the edges highlighted in red, rather than along the edges highlighted in blue. We also note that the clustering coefficient of the green node is equal to 1 (all neighbors are also connected to each other to form a closed triangular motif), while the clustering coefficient of the orange node is $<<1$ (only 3 out of 5 neighbors are also connected to each other). \emph{(B)} In a weighted graph, edges can have different weights. In this example, edges have weights of $3/3=1$, $2/3=0.66$, and $1/3=0.33$. If one wishes to traverse the graph from the orange node to the green node along the shortest path, one would choose to follow the path along the edges with weight equal to unity (stronger weights are equivalent to shorter topological distance). Note also that because the edges are now weighted, neither the orange nor the green nodes has a clustering coefficient equal to unity. \label{cl_binary_weighted}}
\end{center}
\end{figure*}

\begin{figure*}[h]
\begin{center}
\includegraphics[width=0.7\textwidth]{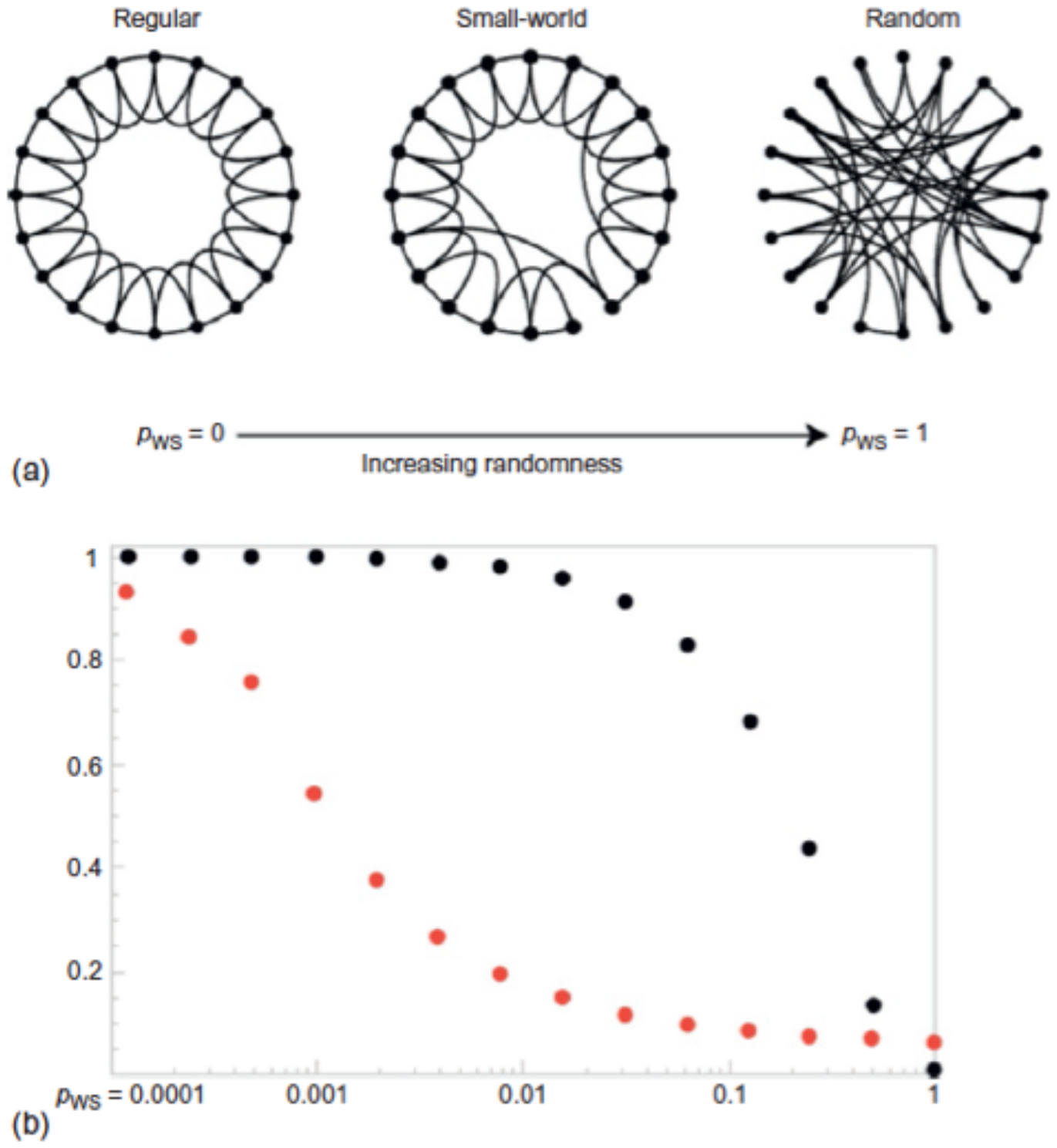}
\caption{ \textbf{The Watts-Strogatz model and the generation of small-world networks.} The canonical model of a small-world network is that described by Duncan Watts and Steve Strogatz in their 1998 paper in Nature \cite{watts1998collective}. The model begins with a regular lattice network in which each node is placed along the circumference of a circle, and is connected to its $k$ nearest neighbors on that circle. Then, with probability $p$, edges are rewired uniformly at random such that (i) at $p=0$, the network is a lattice and (ii) at $p=1$, the network is random. Interestingly, at intermediate values of $p$, the network has so-called ``small-world'' characteristics with significant local clustering (from the lattice model) and short average path-length facilitated by the topological short-cuts created during the random rewiring procedure. Because this architecture can be defined mathematically, small-world graphs have proven fundamental in understanding game theory \cite{li2009largest} and even testing analytical results in subfields of mathematics \cite{konishi2011topology}. Yet, while this work provided a qualitative model of a small-world graph, it did not give a statistic to measure the degree of small-worldness in a particular data set. As a simple scalar measure of ``small-worldness'', Humphries and colleagues defined the \emph{small-world index}, $\sigma$, to be the ratio of the clustering coefficient (normalized by that expected in a random graph) to the average shortest path length (also normalized by that expected in a random graph) \cite{humphries2006brainstem}. The intuition here is that this index should be large (in particular, $\sigma>1$) when the clustering coefficient is much greater than expected in the random graph, and the average shortest path length is comparable to that expected in a random graph. Since this initial definition, other extensions have been proposed and utilized \cite{toppi2012how,telesford2011ubiquity}, building on the same general notions. \label{WS_SW}}
\end{center}
\end{figure*}

Famously, Watts \& Strogatz (1998) \cite{watts1998collective} combined this concept of path length (the minimum number of edges needed to make a connection between nodes) with a measure of topological clustering or cliquishness of edges between nodes (Fig.~\ref{cl_binary_weighted}). More formally, clustering measures the probability that the nodes $j$ and $k$, which are both directly connected to node $i$, are also directly connected to each other; this is equivalent to measuring the proportion of closed triangular 3-node motifs in a network \cite{sporns2004}. Watts \& Strogatz (WS) explored the behaviour of path length and clustering in a simple generative model (henceforth the WS model) (Fig.~\ref{WS_SW}). Starting with a binary lattice network of $N$ nodes each connected to the same number of nearest neighbors, by edges of identical weight (unity), the WS model iteratively re-wires the lattice by randomly deleting an existing edge, between nodes $i$ and $j$, and replacing it by a new edge between node $i$ and any node $k \neq j$. They found that as the probability of random rewiring was incrementally increased from zero, so that the original lattice was progressively randomised, sparsely rewired networks demonstrated both high clustering (like a lattice) and short path length (like a random graph). By analogy to social networks, these algorithmically generated graphs were called small-world networks.  

In addition to introducing this generative model, Watts \& Strogatz also showed how small-worldness could be estimated in naturally occurring networks \cite{watts1998collective}. The hybrid combination of high clustering and short path length that emerged in sparsely re-wired WS networks was proposed as a general quantitative measure of small-worldness (SW) in other networks. It was shown immediately that a nervous system was among the real-world networks that shared the SW pattern of topological organisation. Using data on the synaptic and gap junction connectivity between all $N=302$ neurons in the nervous system of {\em Caenorhabditis elegans} \cite{white1986structure}, a binary undirected graph was constructed representing each neuron as an identical node and each synapse ($\sim 5000$) or gap junction ($\sim 600$) as an identical, unweighted and undirected edge between nodes. This graph of about 5600 edges between 302 nodes was sparsely connected: only about 12\% of the maximum possible number of synaptic connections, $(N^2-N)/2 = 45,451$, actually existed. Compared to a random graph of $N$ nodes, {\em C elegans} had high clustering $\Gamma \sim 5.6$ and short path length $\Lambda \sim 1.18$. Thus the {\em C. elegans} connectome was small-world, in the same quantitative sense as the networks generated by the WS model at low re-wiring probabilities, less than 10\%. But note that does {\em not} necessarily mean that the {\em C. elegans} connectome was biologically generated by the WS algorithm of random rewiring of established connections (axonal projections) between neurons. To put it another way, the WS model can generate SW networks but not all SW networks were generated by a WS model. (And the WS model does not seem like a biologically plausible generative model for brain networks \cite{vertes2012simple,vertes2014generative,betzel2016generative}.)

\subsection*{Small-world brain graphs}

Following the small-world analysis of {\em C elegans}, pioneering topological studies of mammalian cortical networks used databases of tract-tracing experiments to demonstrate that the cat and macaque inter-areal anatomical networks shared similar small-world properties of short path length and high clustering \cite{sporns2004small,hilgetag2004clustered}. The first graph theoretical studies of neuroimaging data demonstrated that large-scale inter-areal networks of functional and structural connectivity in the human brain also had small-world properties \cite{salvador2005neurophysiological,bassett2006adaptive,vaessen2010effect}. These and other seminal discoveries were central to the emergence of connectomics as a major growth point of network neuroscience \cite{sporns2005human}.

About 10 years ago, we reviewed these and other data in support of the idea that the brain is a small world network \cite{bassett2006small}. Here, we aim to take another look at the concept of small-worldness, about one or two decades since it was first formulated quantitatively and applied to brain network analysis at microscopic and macroscopic scales of anatomical resolution. First, we review some of the key questions about small-worldness that have been a focus of work in the period 2006--2016; then we review the technical evidence for small-worldness in high resolution tract-tracing data from the macaque and the mouse; finally, we highlight some likely trends in the further evolution of small-worldness as part of a deeper understanding of the topology of weighted brain graphs.

\section*{What have we (not) learnt since 2006?}

We have learnt a lot about complex topological organisation of nervous systems since 2006, as evidenced by rapid growth in research articles, reviews and citations related to``brain graphs'' and ``connectomes" \cite{bullmore2009complex,bullmore2011brain,pessoa2014understanding}; by the publication of several textbooks \cite{sporns2011networks,fornito2016fundamentals}; and by the recent launch of new specialist journals for network neuroscience \cite{bassett2016network}. This emerging field of brain topology has grown much bigger than the foundational concept of small-worldness. But what have we learnt more specifically about brain small-worldness since 2006, and what do we still have to learn?

\subsection*{Universality}

There is no doubt that small-worldness -- the combination of non-random clustering with near-random path length -- has been very frequently reported across a wide range of neuroscience studies. Small-world topology has been highly replicated across multiple species and scales from structural and functional MRI studies of large scale brain networks in humans to multi-electrode array recordings of cellular networks in cultures \cite{bettencourt2007functional} and intact animals \cite{van2016comparative}. It seems reasonable to conclude that small-worldness is at least very common in network neuroscience; but is it a universal property of nervous systems? Universality is a strong claim and difficult to affirm conclusively. As Popper noted in his philosophy of science by hypothetical refutation \cite{popper1963}, the universal hypothesis that ``all swans are white'' can only be affirmed conclusively by a complete survey of every swan in the world. Whereas it can be immediately and decisively refuted by the observation of a single black swan. Similarly, the claim that {\em all} brains have small-world topology has not yet been (and never will be) affirmed by a complete connectomic mapping of every brain in the world. Some apparent counter-examples of brain networks that do {\em not} have small-world topology have been reported and deserve careful consideration as possible Popperian black swans (see below). However, we can provisionally conclude that enough evidence has amassed to judge that small-worldness is a nearly universal property of nervous systems. Indeed it seems likely that brains are only one of a large ``universality class'' of small-world networks comprising also many other non-neural or non-biological complex systems. Such near-universality of small-worldness, or any other brain network parameter, has a number of implications.

First, near-universality implies {\em self-similarity}. If the macro-scale inter-areal network of the human brain is small-world, as is the micro-scale inter-neuronal network of the worm or the fly, then we should expect also that the micro-scale inter-neuronal network of the human brain is small-world. Self-similarity of small-worldness would be indexed by scale invariance of network path length and clustering parameters as the anatomical resolution ``zooms in'' from macro- to micro-scales. Although there is abundant evidence for scaling, fractal or self-similar statistics in many aspects of brain network topology \cite{bullmore2009generic,bassett2010efficient,klimm2014resolving}, experimental data do not yet exist that could support a multi-scale, macro-to-micro analysis of small-worldness (and other network properties) in the same (human or mammalian) nervous system \cite{bassett2013multiscale}.

Second, near-universality suggests some very general selection pressures might be operative on the {\em evolution and development} of nervous systems across scales and species. This line of thinking has led to the formulation of generative models that can simulate brain networks by some probabilistic growth rule or genetic algorithm. It has been found that simple generative models, that add edges to a network based on the spatial distance and the topological relationships between nodes, can recapitulate small-worldness and many other properties of the connectome on the basis of two (spatial and topological) parameters \cite{vertes2012simple,vertes2014generative,betzel2016generative}. This serves as a reminder that the network phenotype of small-worldness can be generated by many different mechanisms and the biological mechanisms controlling formation of small-world properties in brain networks currently remain unknown.

Third, and from a somewhat more controversial perspective, universality might seem tantamount to {\em triviality}. If the brain is everywhere small-world, and so are almost all other complex systems in real-life \cite{bassett2006small,bullmore2009generic,gaiteri2014beyond,moslonka2011networks,sizemore2016classification} (for a few exceptions, see \cite{koschutzki2010structural}), then what is the small-worldness of the brain telling us that's of any interest specifically to neuroscience? There are two main answers to this important question, as we discuss in more detail below: (i) studies have recently succeeded in linking network topological metrics to biological concepts, like wiring cost \cite{bullmore2012economy,bassett2010efficient,bassett2011conserved,rubinov2015wiring}, and to biological phenotypes, like neuronal density \cite{acimovic2015effects,vandenheuvel2015bridging} or gene expression \cite{fulcher2016transcriptional,vertes2016philtrans,whitaker2016pnas}; and (ii) small-worldness is not the whole story of brain network organisation \cite{wang2016brain}.

\subsection*{Economical small-world networks}

At the risk of stating the obvious, small-worldness is a purely topological quantity that tells us nothing about the physical layout of the nodes or edges that constitute the graph \cite{bassett2010efficient,pessoa2014understanding}. However, it is equally obvious that brain networks are embedded in anatomical space \cite{klimm2014resolving,lohse2014resolving,bassett2011conserved,betzel2016modular}. Somehow the abstract, dimensionless topology of small-worldness must be reconciled to the anatomy of the brain. It turns out that the small-world topology of brain networks is (almost) always {\em economically} embedded in physical space \cite{bullmore2012economy,kaiser2006nonoptimal}.

For both clustering and path length, the two topological metrics combined in the hybrid small-world estimator, there is a strong relationship with brain anatomical space \cite{bassett2010efficient,bassett2011conserved,rubinov2015wiring}. The edges between clustered nodes tend to be shorter distance whereas the edges that mediate topological short cuts tend to traverse longer anatomical distances. Interpreting the Euclidean distance between brain regional nodes or neurons as a proxy for the wiring cost, i.e., the total biological cost of building a physical connection and maintaining communication between nodes, it has been argued that the brain is an economical small-world network \cite{latora2001efficient,bullmore2012economy}. Economical in this sense does not simply mean parsimonious or cheap; it is more closely related to the common-sense notion of ``value for money''. Topologically clustered nodes are anatomically co-located and thereby nearly minimise wiring cost. But small-world brain networks are not naturally lattices and if they are computationally rewired strictly to minimise wiring cost then brain networks are topologically penalised, losing integrative capacity indexed by increased characteristic path length and thus reduced small-worldness scalar $\sigma$.

The economical idea is that brain networks have been selected by the competition between a pressure to minimise biological cost \emph{versus} a pressure to maximise topological integration. More formally,
\begin{equation}
P_{i,j} \sim f(d_{i,j}) f(k_{i,j}),
\end{equation}
the probability of a connection between nodes $i$ and $j$, $P_{i,j}$, is a product of: a function of the physical distance in mm between nodes $d_{i,j}$ - often used as a proxy for wiring cost; and a function of the topological relationship between nodes - $k_{i,j}$. 

Typically the functions of cost and topology are each parameterised by a single parameter, for example, simple exponential and power law functions. Several variants of this approach have been published, exploring a range of different topological relationships $k_{i,j}$ between nodes, for example, clustering and homophily \cite{vertes2012simple,vertes2014generative,betzel2016generative}. Economical models can generally reproduce the small world properties of brain networks quite realistically: clustering and path length are both increased as a function of the cost parameter \cite{avena2014using}. In other words, as the cost penalty becomes the dominant factor predicting the probability of a connection between nodes, economical models generate increasingly lattice-like networks, with strong spatial and topological clustering of connected nodes, approximating in the limit the minimal cost configuration of the network. The emergence of more integrative network features -- such as hubs mediating many inter-modular connections -- typically depends on some degree of relaxation of the cost penalty (reduced distance parameter) relative to the parameter controlling the importance of (integrative) topological relationships between nodes in predicting their connectivity. Thus small world networks can be generated by economical models for a certain range of the two parameters controlling the competitive factors of (wiring) cost and (topological) value.

\subsection*{Small-worldness is not the whole story}

Before getting further into the details of small-worldness, as we do below in relation to recent tract-tracing results, it is important to acknowledge that the specific metrics of path length $\Lambda$ and clustering $\Gamma$ introduced by Watts \& Strogatz \cite{watts1998collective}, and the small-worldness scalar derived from them $\sigma = {\Gamma}/{\Lambda}$ \cite{humphries2006brainstem}, are a few global topological metrics that have been of central importance to the growth of complex network science generally. But more than 15 years after the first discovery of small world properties in brain networks, the field of connectomics now extends into many other areas of topological analysis. There is much important recent work on topological properties like degree distribution and hubness \cite{achard2006resilient}, modularity \cite{simon1962architecture,meunier2009age,chen2008revealing,sporns2016modular,bassett2011dynamic,mattar2015functional,stoop2013beyond}, core/periphery organisation \cite{bassett2013task,senden2014rich,van2011rich}, controllability \cite{gu2015controllability,muldoon2016stimulation,betzel2016optimally} and navigability \cite{gulyas2015navigable} that are not simply related to small-worldness. Outside neuroscience there continues to be strong growth in the more general field of network science \cite{barabasi2016book}. It is nothing like a complete description of the brain to say it is small-world; we now turn to a more technical discussion of the evidence for small-worldness as a common property of nervous systems.

\section*{Challenges to small-worldness}

\begin{figure*}[h]
\begin{center}
\includegraphics[width=0.9\textwidth]{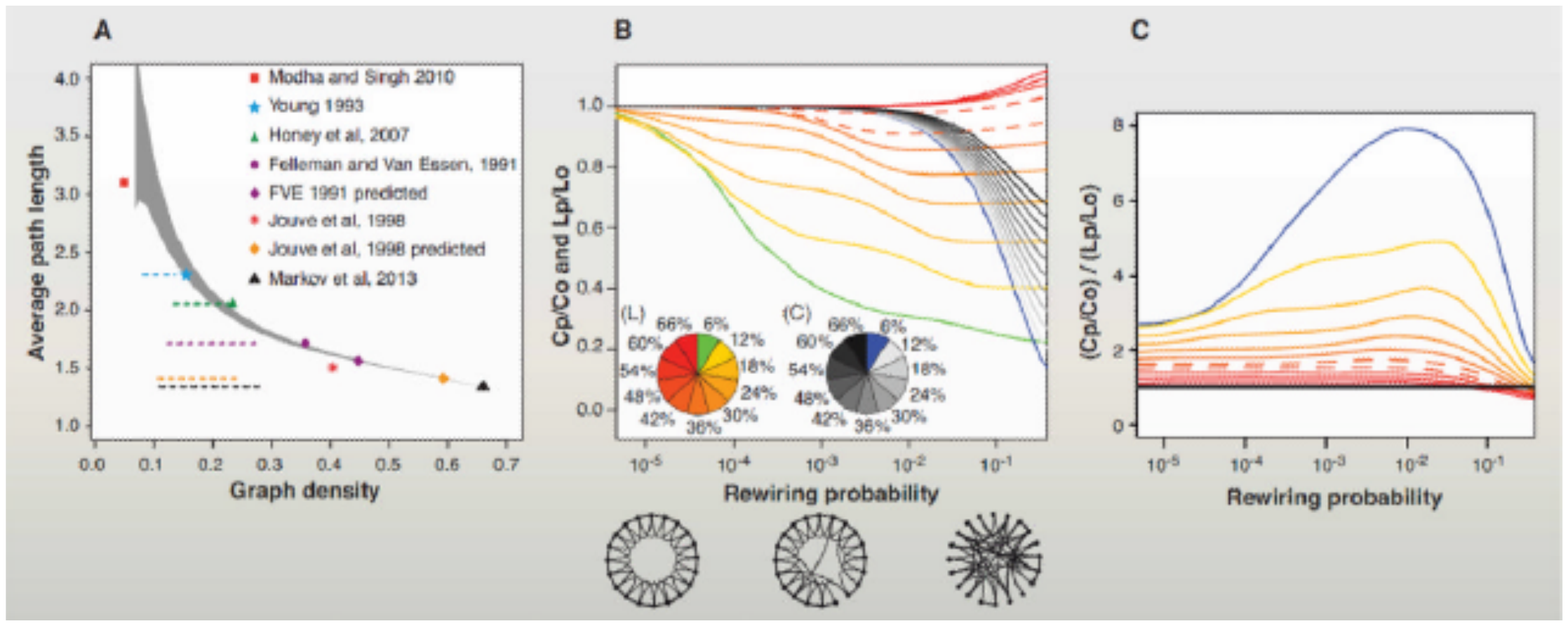}
\caption{\textbf{High density of the macaque cortical graph excludes sparse small world architecture} \emph{(A)} Comparison of the average shortest path length and density of the macaque cortical graph from \cite{markov2013cortical} with the graphs of previous studies. Sequential removal of weak connections causes an increase in the path length. Black triangle: macaque cortical graph from \cite{markov2013cortical}; gray area: 95\% confidence interval following random removal of connections from the macaque cortical graph from \cite{markov2013cortical}. Modha and Singh 2010: \cite{modha2010network}; Young 1993: \cite{young1993organization}; Honey et al., 2007: \cite{honey2007network}; Felleman and Van Essen 1991: \cite{felleman1991distributed}; Jouve et al., 1998: \cite{jouve1998mathematical}; Markov et al., 2012: \cite{markov2014weighted}. “Jouve et al., 1998 predicted” indicates values of the graph inferred using the published algorithm \cite{jouve1998mathematical}. \emph{(B)} Effect of density on Watts and Strogatz's formalization of a small world network. Clustering and path length variations generated by edge rewiring with probability range indicated on the $x$-axis applied to regular lattices of increasingly higher densities. The pie charts show graph density encoded via colors for path length ($L$) and clustering coefficient ($C$). The $y$-axis indicates the path length ratio ($L_{p}$/$L_{o}$) and clustering ratio ($C_{p}$/$C_{o}$) of the randomly rewired network, where $L_{o}$ and $C_{o}$ are the path length ($L_{o}$) and clustering ($C_{o}$) of the regular lattice, respectively. The variables $L_{p}$ and $C_{p}$ are the same quantities measured for the network rewired with probability $p$. Hence, for each density value indicated in the $L$ and $C$ pie charts, the corresponding $L_{p}$/$L_{o}$ and $C_{p}$/$C_{o}$ curves can be identified. Three diagrams below the $x$-axis indicate the lattice (\emph{left}), sparsely rewired (\emph{middle}), and the randomized (\emph{right}) networks. \emph{(C)} The small-world coefficient $\sigma$ \cite{humphries2006brainstem} corresponding to each lattice rewiring. Color code is the same as in panel \emph{(B)}. Dashed lines in \emph{(B)} and \emph{(C)} indicate 42\% and 48\% density levels. Reproduced with permission from \cite{markov2013cortical}. \label{kennedy}}
\end{center}
\end{figure*}

About 3--4 years ago, an important series of papers began to be published that could be regarded as ``black swans'' refuting the general importance of small-worldness in an understanding of brain networks \cite{markov2012cerebcortex,markov2013cortical,song2014spatial,knoblaugh2016}:
\begin{quote}
Previous studies of low density inter-areal graphs and apparent small-world properties are challenged by data that reveal high-density cortical graphs in which economy of connections is achieved by weight heterogeneity and distance-weight correlations. \cite{markov2013cortical}
\end{quote}
\begin{quote}
Recent connectomic tract tracing reveals that, contrary to what was previously thought, the cortical inter-areal network has high density. This finding leads to a necessary revision of the relevance of some of the graph theoretical notions, such as the small world property..., that have been claimed to characterise the inter-areal cortical network. \cite{knoblaugh2016}
\end{quote}
These remarks carried weight because they were based on sophisticated and highly sensitive measurements of mammalian cortical connectivity (Fig.~\ref{kennedy}). In each one of multiple carefully standardised experiments in the macaque monkey, a fluorescent tracer was injected into a (target) cortical region where it was taken up by synaptic terminals and actively transported to the cell bodies of neurons projecting to the target region. When the animal's brain was subsequently examined microscopically, the retrograde transport of the tracer from the injection site resulted in a fluorescent signal in the (source) regions of cortex that were directly connected to the target region. The basic technology of anatomical tract-tracing had been used by neuroanatomists since the late 20th century; but in the first decades of the 21st century it was possible to increase the scale and precision of the measurements dramatically, enabling the construction of connectivity matrices that summarised the strength or weight of axonal projections between a large number of cortical areas. These next-generation tract-tracing data thus represented a new standard of knowledge about mammalian cortical connectivity, that was more continuously quantified than the binary or ordinal rating of connectivity from traditional tract-tracing experiments \cite{stephan2001cocomac}, and much less ambiguously related to the cellular substrates of brain networks than the statistical measures of functional connectivity \cite{achard2006resilient,zhang2016choosing} and structural covariance \cite{bassett2008hierarchical,alexander2013imaging} used to build graphs from human neuroimaging data. It is clearly important to understand in some detail how the topology of brain networks can be modelled in contemporary tract-tracing data from the macaque (and subsequently the mouse \cite{oh2014mesoscale,rubinov2015wiring}) and what these results tell us about the small-worldness of brain networks.

\subsection*{Binary graphs}

In general, a node represents a component of a system and an edge represents a connection or interaction between two nodes. Mathematically, we can capture these ideas with a graph $\mc G = (\mc V, \mc E)$ composed of a node set $\mc V$ and an edge set $\mc E$ \cite{bollobas1979graph,bollobas1985random}. We store this information in an association or weight matrix $\mathbf{W}$, whose $ij^{th}$ element indicates the strength or weight $w_{i,j}$ of the edge between node $i$ and node $j$. A simple way of building a graph from such an association matrix is to apply a threshold $\tau$ to each element of the matrix, such that if $w_{i,j} \geq \tau$ then an edge is drawn between the corresponding nodes, but if $w_{i,j} < \tau$ no edge is drawn \cite{achard2006resilient}. This thresholding operation thus binarizes the weight matrix and converts the continuously variable edge weights to either 1 (supra-threshold) or 0 (sub-threshold). It was on this basis that almost all brain graphs were constructed in the 15 years or so following the seminal small-world analysis of a binary graph representing the cellular connectome of {\em C. elegans} \cite{watts1998collective}. Most of the neuroimaging evidence for small-worldness in human brain networks, for example, is based on analysis of binary graphs constructed by thresholding a correlation coefficient or equivalent estimator of the weight of functional or structural connectivity or structural covariance between regions $i$ and $j$ \cite{wijk2010comparing}. It is well recognised that construction of binary graphs represents an extreme simplification of brain networks; indeed a binary undirected graph of homogenous nodes is as simple as it gets in graph theory \cite{bassett2012altered}. However, this approach has historically been preferred in neuroimaging because of limited signal-to-noise ratio in the data \cite{achard2006resilient}.

By varying the threshold $\tau$ used to construct a binary graph from a continuous weight matrix, the connection density of the network is made denser or sparser. If the threshold is low and many weak weights are added to the graph as edges then the connection density will increase; if the threshold is high and only the strongest weights are represented as edges, then the connection density will decrease. The connection density $D$ is quantified by the number of edges $E$ in the graph as a proportion of the total number of edges in a fully connected network of the same number of nodes $N$:
\begin{equation}
D   =   \frac{E}{N^2-N/2}
\end{equation} Often, this proportion is translated into a percentage. In many neuroimaging studies, the threshold is set to a large value to control for the high levels of noise in MRI data, resulting in connection densities in the range $5-30\%$ \cite{lynall2010functional}. In many of the first generation tract tracing studies, the connectivity data were collected on a binary or ordinal scale, and not all possible connections had been been experimentally measured, so these data were naturally modelled as binary graphs with connection densities $\sim 30\%$, a value that was constrained by the completeness and quality of the data \cite{bassett2006small}.

The small-world topology of a binary brain graph is defined by estimating two parameters in the data, path length $L$ and clustering $C$ (Fig.~\ref{cl_binary_weighted}A), and comparing each of these observed parameters to their distributions under a specified null model \cite{humphries2006brainstem}. More specifically,
\begin{equation}
L  =  \frac{1}{N} \sum l_{i,j}
\end{equation}
\noindent is the global or characteristic path length, where $l_{i,j}$ is the shortest path (geodesic) between nodes $i$ and $j$; and
\begin{equation}
C =  \frac{1}{N} \sum c_{i,j}
\end{equation}
\noindent is the global clustering coefficient, where $c_{i,j}$ is the number of closed triangular motifs including node $i$. Each of these parameters is normalised by its value in a binary graph representing the null hypothesis. For example, if the null hypothesis is that clustering of brain networks $C_{brain}$ is no different from the clustering of a random graph, then it is reasonable to generate an Erd\"os-Reny\'i graph for $N$ nodes and $D$ connection density, measure the clustering coefficient in the random graph $C_{random}$, and use the ratio between brain and random graph clustering coefficients as a test statistic for non-random clustering. We note that there are many other possible ways in which a null model could be sampled, besides using the classical Erd\"os-Reny\'i model, and this is an active area of methodological research \cite{muldoon2016small,betzel2016modular}. However, in general one can define the normalized clustering coefficient as
\begin{equation}
\Gamma = \frac{C_{brain}}{C_{random}}.
\end{equation}
Likewise, the path length of the brain graph can be normalised by its value in a comparable random graph
\begin{equation}
 \Lambda = \frac{L_{brain}}{L_{random}}.
\end{equation}
A small-worldness scalar can then be simply defined as
\begin{equation}
\sigma = \frac{\Gamma}{\Lambda}.
\end{equation}
With these definitions, small-world networks will have $\sigma > 1$, $\Gamma > 1$ and $\Lambda \sim 1$ \cite{humphries2006brainstem}.

\subsection*{Weighted graphs}

Although binary graph analysis has predominated to date in analysis of brain networks, this certainly does not represent the methodological limit of graph theory for connectomics. For example, provided the data are of sufficient quality, there is no need to threshold the weight matrix to estimate topological properties like clustering, path length and small-worldness. Indeed, while the binarization procedure was common in early applications of graph theory to neural data \cite{wijk2010comparing}, it remains fundamentally agnostic to architectural principles that may be encoded in edge weights \cite{rubinov2011weight}. This realization has more generally motivated the field to develop methods that remain sensitive to the patterns of weights on the edges \cite{ginestet2011brain}, and to the topologies present in weak \emph{versus} strong weights \cite{rubinov2011weight}. These efforts have included the development of alternative thresholding schemes \cite{bassett2012altered,lohse2014resolving} and fully weighted graph analysis \cite{rubinov2011weight,bassett2011dynamic}.

The mathematical tools exist to estimate and simulate the topological properties of weighted networks, and analysis of weighted networks is akin to studying the geometry of the graph, rather than simply its topology  \cite{bassett2012influence,bassett2013task}. For example, weighted analogues of binary metrics of clustering, path length and small-worldness can be defined formally (Fig.~\ref{cl_binary_weighted}B). First, the weighted clustering coefficient of node $i$ can be defined as

\begin{equation}
C_{\mathrm{weighted}} = \frac{1}{k_i(k_i-1)}\sum\limits_{j,k}(\hat{w}_{ij}\hat{w}_{jk}\hat{w}_{ik})^{1/3},
\label{c_weighted}
\end{equation}
\noindent where $k_i$ is the number of edges connected to node $i$, or degree of node $i$ \cite{Onnela:2005de} (but see also \cite{Barrat:2004bk,Zhang:2005er} for other similar definitions). The weighted path length can be defined as
\begin{equation}
L_{\mathrm{weighted}}=\frac{1}{N(N-1)}\sum_{i\ne j}\delta_{ij},
\label{l_weighted}
\end{equation}

\noindent where the topological distance between two nodes is given by $\delta_{ij}=1/w_{ij}$  \cite{Newman:2001kc}. These two statistics can be combined to construct a weighted metric of small-worldness \cite{bolanos2013weighted}:
\begin{equation}
\sigma_{weighted}= \frac{\Gamma_{weighted}}{\Lambda_{weighted}}.
\end{equation} 
 With these definitions, small-world networks will have  $\Gamma_{\mathrm{weighted}} > 1$, $\Lambda_{\mathrm{weighted}} \sim 1$ and $\sigma_{\mathrm{weighted}} > 1$, \cite{humphries2006brainstem}.

\subsection*{The small-world propensity}

\begin{figure*}[h]
\begin{center}
\includegraphics[width=0.6\textwidth]{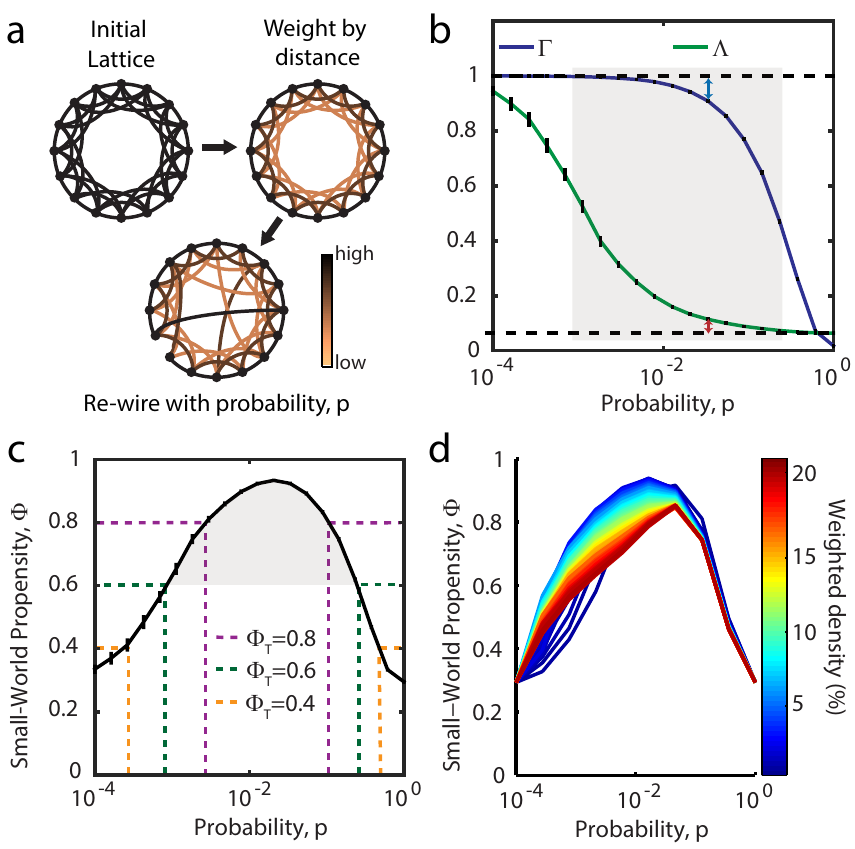}
\caption{ \textbf{Small-world propensity in weighted networks.} Here we illustrate an example of a generative small-world model, and its utility in estimating an empirical network's small-world propensity. \emph{(A)} We can extend the concept of a Watts-Strogatz model to weighted graphs by first building a lattice in which the edges are weighted by distance such that edges between spatially neighbouring nodes have more strongly weighted than edges between spatially distant nodes. These edge weights can then be rewired with a probability, $P$, to create a weighted small-world network. \emph{(B)} Weighted clustering coefficient and weighted path length can be estimated as a function of the rewiring parameter, $P$, and used to derive the small-world propensity of the graph compared to random and lattice benchmarks (Eq \ref{eq:SWP}). \emph{(C)} Weighted small-world propensity calculated for the same network as in panel \emph{(B)}. Error bars represent the standard error of the mean calculated over 50 simulations, and the shaded regions represent the range denoted as small-world. \emph{(D)} Weighted small-world propensity as a function of network density for a graph of 1000 nodes. Reproduced with permission from \cite{muldoon2016small}. \label{smallworldpropensity}}
\end{center}
\end{figure*}

There are several important limitations to the definitions of small-worldness described in the previous sections. First, the small-world scalar $\sigma$ (whether binary or weighted) can be greater than 1 even in cases when the normalized path-length is much greater than one; because it is defined as a ratio, if $\Gamma >> \Lambda > 1$, the scalar $\sigma > 1$. This means that a small-world network will always have $\sigma > 1$, but not all networks with $\sigma > 1$ will be small-world.   Second, the measure is strongly driven by the density of the graph, and denser networks will more naturally have smaller values of $\sigma$ even if they are in fact generated from an identical small-world model. To address these and other limitations, Muldoon and colleagues recently developed a metric called the \emph{small-world propensity}. Specifically, the small-world propensity, $\phi$, reflects the deviation of a network's clustering coefficient, $C_{brain}$, and characteristic path length, $L_{brain}$, from both lattice ($C_{lattice}$, $L_{lattice}$) and random ($C_{random}$, $L_{random}$) networks constructed with the same number of nodes and the same degree distribution:

\begin{equation}
\label{eq:SWP}
\phi = 1-\sqrt{\frac{\Delta_{C}^{2}+\Delta_{L}^{2}}{2}},
\end{equation}
where
\begin{equation}
\Delta_{C}=\frac{C_{lattice}-C_{brain}}{C_{lattice}-C_{random}}
\end{equation}
and
\begin{equation}
\Delta_{L}=\frac{L_{brain}-L_{random}}{L_{lattice}-L_{random}}.
\end{equation}

\noindent The ratio $\Delta_{C/L}$ represents the fractional deviation of the metric ($C_{brain}$ or $L_{brain}$) from its respective null model (a lattice or random network). This quantity can be calculated for binary networks (using binary definitions of clustering and path length) or for weighted networks (using weighted definitions of clustering and path-length). Networks are considered small-world if they have small-world propensity $0.4 < \phi \leq 1$.  However, this metric should be viewed as a continuous metric of small-worldness rather than a hard threshold \cite{muldoon2016small}.

Importantly, the small-world propensity overcomes several limitations of previous scalar definitions of small-worldness \cite{muldoon2016small}. First, it can incorporate weighted estimates of both the clustering coefficient and path-length, thus being generally applicable to any neural data that can be represented as a weighted network. Second, it is density-independent, meaning that it can be used to compare the relative small-worldness between two networks that have very different densities from one another. Third, the metric is informed by spatially-constrained null models \cite{expert2011uncovering,bassett2015extraction,papadopoulos2016evolution} in which nodes have physical locations and the edges that correspond to the smallest Euclidean distance between nodes are assigned the highest weights \cite{barthelemy2011spatial} (Fig \ref{smallworldpropensity}).

\section*{21st century tract-tracing}

The scale and quality of contemporary tract-tracing data, in both the macaque and the mouse, represents a quantitative change in terms of sensitivity in detecting anatomical connections, or axonal projections, between cortical areas. Using retrograde tracer experiments it has proven possible to demonstrate reliably that pairs of regions in the macaque brain may be connected by one or a few axonal projections. Likewise anterograde tracer experiments in the mouse have demonstrated that the minimal detectable weight of connectivity between cortical regions, that just exceeds the noise threshold, is equivalent to the projection of one or a few axons \cite{ypma2016}. This high sensitivity has led immediately to the recognition of a large number of weak and previously unreported axonal connections. In the macaque, it was estimated that 36\% of connections identified by contemporary tract tracing were so-called new found projections (NFP) that had not been described in the prior literature \cite{markov2014weighted}. The existence of so many weak connections is reflected in the log normal distributions of connectivity weight, ranging over 5--6 orders of magnitude, in both the macaque and the mouse \cite{oh2014mesoscale,ercsey2013predictive}. In short, tract-tracing can now resolve connections approximately equivalent to a single axonal projection and approximately a million times weaker than the strongest anatomical connections or white matter tracts.

How can we use graph theory to model the network organisation of such highly sensitive, highly variable data? Perhaps the simplest approach, borrowing from prior studies of less high quality datasets, is to apply a threshold and convert the log-normal weight matrix into a binary adjacency matrix. If the threshold is defined by the noise distribution of the measurements then it will be very close to zero for these sophisticated experiments, and correspondingly the connection density of the binary graph will be high. In the macaque, the connection density of a binary graph of 29 visual cortical areas was estimated to be 66\% \cite{markov2013cortical}, considerably higher than historical estimates in the range $25\%-45\%$ \cite{felleman1991distributed}. In the mouse, the connection density of a binary graph of 308 areas of the whole cortex was estimated to be 53\% \cite{rubinov2015wiring}.

In other words, the binary graphs generated from 21st century tract-tracing data are about twice as dense as the much sparser networks derived from human neuroimaging and 20th century tract-tracing. They are also considerably denser than brain networks constructed at a finer grained (ultimately cellular) resolution. For example, the connection density of the {\em C elegans} nervous system, which is still the only completely mapped synaptic connectome, is about 12\%. It is easy to see that the connection density of a binary network depends on the number of neurons comprising each node. In the limit, if the nervous system is parcellated into two large nodes the connection density will certainly be 100\%; as the same system is parcellated into a larger number of smaller nodes its connection density will monotonically decrease \cite{bassett2011conserved,zaleksy2010whole}. Thus the current interval estimate of mammalian cortical connection density $\sim 55-65\%$ is conditional both on the anatomical resolution of the parcellation scheme used to define the nodes and the sensitivity of the tract-tracing methods used to estimate the weights of the edges.

\begin{figure*}[h]
\begin{center}
\includegraphics[width=0.7\textwidth]{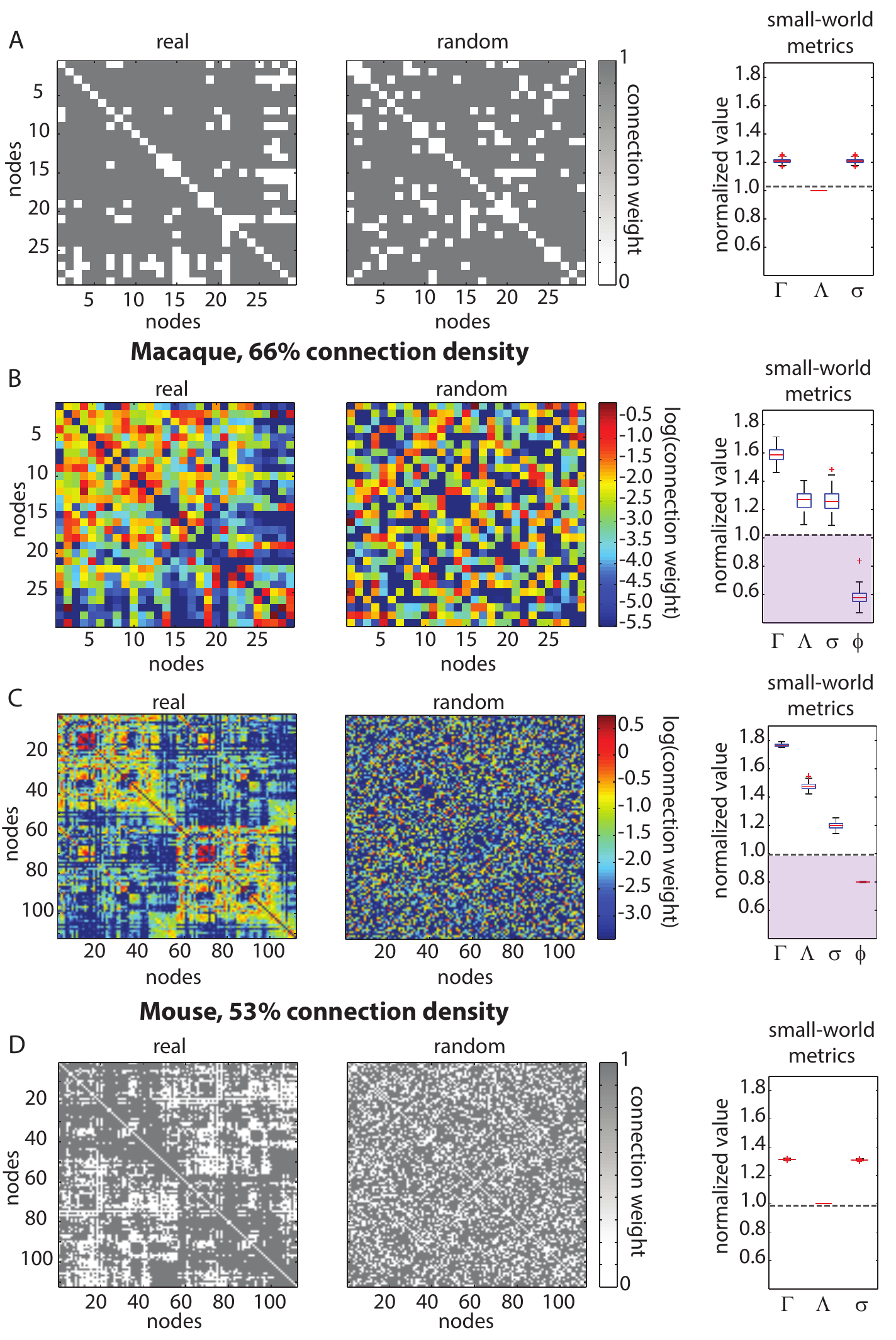}
\caption{ \textbf{Binary and weighted small-worldness in mouse and macaque connectomes.} For the macaque connectome reported in \cite{markov2013cortical}, we show \emph{(A)} the binary network, a random graph of the same size and density, and the estimated small-world parameters $\Gamma$ (normalized clustering coefficient), $\Lambda$ (normalized path length), $\sigma$ (classical small-world scalar) and $\phi$ (small world propensity). In panel \emph{(B)} we show a weighted network analysis for the same data. For the mouse connectome reported in \cite{rubinov2015wiring}, we show \emph{(C)} the weighted network, a random graph of the same size and density, and the estimated small-world parameters $\Gamma$ (normalized clustering coefficient), $\Lambda$ (normalized path length), $\sigma$ (classical small-world scalar) and $\phi$ (small world propensity). In panel \emph{(D)} we show a binary network analysis for the same data. In the boxplots, the gray dotted line shows the threshold value of $\sigma = 1$, and the purple area shows the range of values of $0.4 < \phi \leq 1$ in which a network is considered small-world. \label{fig_sw_mouse_macaque}}
\end{center}
\end{figure*}

\subsection*{Small-worldness of binary tract-tracing networks}

Having constructed a high density binary graph from tract-tracing data on mammalian cortex, it is straightforward to estimate its clustering and path length, using the same metrics as for sparser binary graphs. However, simply because there is a larger number of connections in the denser network, its clustering will be considerably higher (there will be more closed triangular motifs) and its path length will be shorter (there will be more direct, pair-wise connections) than a sparser network. Indeed the clustering and path length of any binary graph at $60\%$ connection density will be close to the maximal clustering and minimal path length of a fully connected graph; and therefore the clustering and path length of a 60\% dense brain network will be very similar to the clustering and path length of a 60\% random network \cite{bassett2009cognitive}. 

This means that when clustering and path length in brain networks are normalised by their corresponding values in equally dense random networks, the scaled metrics $\Gamma$ and $\Lambda$ will both be close to 1, and the small-world scalar $\sigma$ will be close to its critical value of 1 \cite{markov2013cortical}. For the macaque, at 66\% connection density, $\Gamma =1.21 \pm 0.014$, $\Lambda=1.00 \pm 0.000$, and $\sigma =1.21\pm0.014$; for the mouse, at 53\% connection density, $\Gamma =1.31 \pm 0.004$, $\Lambda=1.00 \pm 0.000$, and $\sigma =1.31 \pm 0.004$ (all given in mean $\pm$ standard deviation; Fig.~\ref{fig_sw_mouse_macaque}A,C; Table~\ref{tab1}). Since small-worldness has been traditionally defined as $\sigma > 1$, these results suggest that dense binary graphs constructed from tract tracing data are small-world, although the macaque is more similar to a random network than the mouse.  

\begin{table*}[ht]
\hfill{}
\begin{tabular}{ |l|l|l|l|l|l|}
\hline
~& Macaque & ~ & ~ & Mouse & ~\\
\hline
~ & Binary & Weighted & ~ & Binary & Weighted \\
$\Gamma$ & $1.21 \pm 0.014$ & $1.59\pm0.007$ & ~ & $1.31 \pm 0.004$ & $1.76\pm0.009$ \\
$\Lambda$ & $1.00 \pm 0.000$ & $1.27 \pm 0.057$ & ~ & $1.00 \pm 0.000$ & $1.47\pm0.021$ \\
$\sigma$ & $1.21\pm0.014$ & $1.25 \pm 0.071$ & ~ & $1.31 \pm 0.004$ & $1.20\pm0.019$ \\
$\phi$ & N/A & $0.574\pm0.041$ & ~ & N/A & $0.800\pm0.002$ \\
\hline
\end{tabular}
\hfill{}
\caption{\textbf{Small-world metrics} For the macaque and mouse connectomes, we show the mean and standard deviation of the normalized clustering coefficient ($\Gamma$), the normalized path length ($\Lambda$), the small-world index $\sigma$, and the small-world propensity $\phi$ for both binary and weighted graphs. }
\label{tab1}
\end{table*}

These results do not look like a ``black swan'' that refutes universal claims that the brain always embodies small-world network topology. Nor do they undermine the credibility of previous studies demonstrating small-worldness in sparser brain graphs. However, our view is that binary graph models are very unlikely to be an optimal strategy for network analysis of tract-tracing data, because they fail to take account of the extraordinary range of connectivity weights, distributed log normally over 6 orders of magnitude, that has been discovered in mammalian cortical networks \cite{ercsey2013predictive}. The weakest connection between cortical areas is about a million times less weighted than the strongest connection: does it really make sense to set all these weights equivalently to 1 as edges in a binary graph? To ask the question is to answer it.

\subsection*{Small-worldness of weighted tract-tracing networks}

A weighted small-world analysis is easily done for these data (Fig.~\ref{fig_sw_mouse_macaque}B,D). The weighted clustering and weighted path length metrics (Eq.~\ref{c_weighted} and Eq.~\ref{l_weighted}) are estimated directly from the weight matrices, and the ratio of weighted clustering to weighted path length is the scalar summary of weighted small-worldness $\sigma_{\mathrm{weighted}} > 1$. In Fig.~\ref{fig_sw_mouse_macaque}, we directly compare binary and weighted graph theoretical results for the mouse \cite{oh2014mesoscale,rubinov2015wiring} and macaque \cite{markov2013cortical} connectomes. Compared to the results of binary graph analysis, both mouse and macaque networks have increased clustering for the weighted graph analysis, and $\sigma$ is increased for the macaque (see Table~\ref{tab1}). 

\begin{figure*}[h]
\begin{center}
\includegraphics[width=1\textwidth]{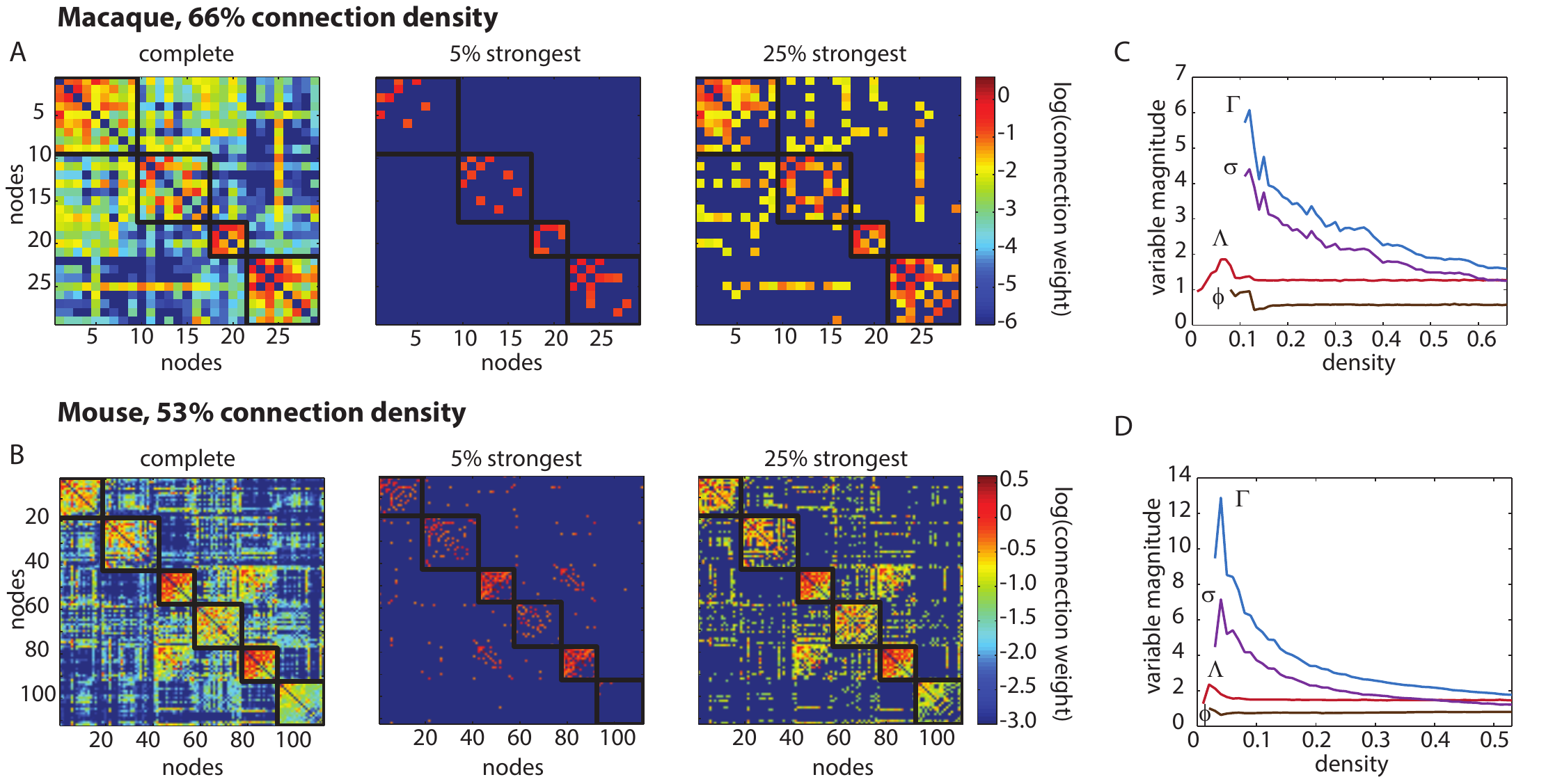}
\caption{ \textbf{Dependence of small-world characteristics on network density.} \emph{(A)} Macaque and \emph{(B)} mouse connectivity matrices in their natural state (\emph{left}), as well as after thresholding to retain the 5\% strongest (\emph{middle}) or 25\% strongest (\emph{right}) connections. Weighted small-world metrics including the normalized clustering coefficient ($\Gamma$), normalized path-length ($\Lambda$), small-world index ($\sigma$), and small-world propensity ($\phi$) as a function of network density for the \emph{(C)} macaque and \emph{(D)} mouse connectivity matrices. \label{fig_sw_mouse_macaque_density}}
\end{center}
\end{figure*}

The weighted graph of the mouse connectome is similarly small-world compared to the weighted macaque graph, as measured by $\sigma$, but is significantly more small-world as measured by the small-world propensity $\phi$. However, classical estimates of small-worldness may depend in a non-trivial way on the density of the graph. This relationship becomes obvious if we estimate the topology of both weighted graphs as a function of connection density  (Fig.~\ref{fig_sw_mouse_macaque_density}). The classical small-world scalar $\sigma$ is greatest when it is estimated for a sparse graph comprising less than 20-30\% of the most strongly connected edges, and decreases progressively as the graph becomes denser.  This might suggest that the macaque connectome seems less small-world than the mouse simply because it is denser. However, the small world propensity $\phi$ has the useful property that it is independent of network density and it is significantly greater, indicating more small-worldness,  for the mouse than the macaque. This could be related to differences between the datasets in number of cortical areas and completeness of cortical coverage: the macaque dataset comprises fewer nodes of mostly visual cortex than the larger number of nodes across the whole mouse cortex.

\subsection*{Weighted small-worldness and the role of edge weights}

\begin{figure*}[t]
\begin{center}
\includegraphics[width=1\textwidth]{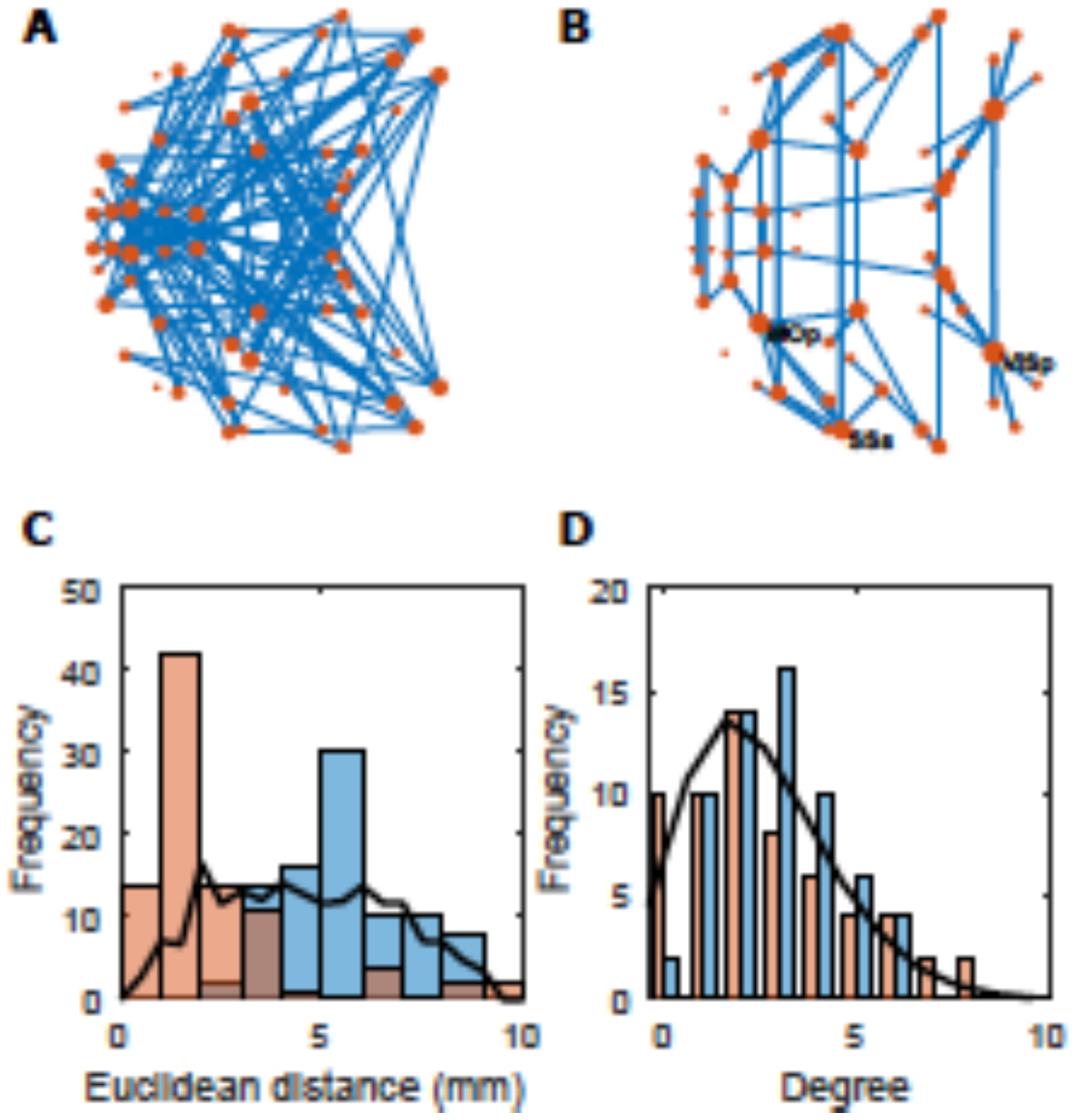}
\caption{ \textbf{The existence of weak links and their topology in the mouse connectome.} Here we show the properties of the 5\% strongest and 5\% weakest edges of the mouse cortical network. \emph{(A,B)} Axial view of the mouse cortical network, red dots represent brain regions, blue lines represent the connections between them. Drawn are the \emph{(A)} 5\% weakest or \emph{(B)} 5\% strongest edges. Dot size corresponds to degree, the total number of incoming and outgoing edges connected to a node. In \emph{(B)}, the three nodes with highest degree have been labeled: VISp, primary visual area; MOp, primary motor area; SSs, supplemental somatosensory area. The strong connections are spatially organized, mainly connecting spatially adjacent or contralaterally homologous regions. The weak connections span longer distances and are topologically more random than the strongest connections. \emph{(C)} The distance distributions for (blue) the 5\% weakest edges, (red) the 5\% strongest edges, and (black) a random graph of the same size and connection density. \emph{(D)} The degree distributions for the weakest and strongest connections of the mouse connectome, and a comparable random graph, color-coded as
in panel \emph{(C)}. Reproduced with permission from \cite{ypma2016}. \label{WeakLinks}}
\end{center}
\end{figure*}

Why does a weighted graph analysis provide stronger evidence for non-random clustering than a binary graph analysis applied to the same tract-tracing data?
The most strongly weighted connections generally span the shortest physical distances between cortical areas \cite{klimm2014resolving,ercsey2013predictive,rubinov2015wiring}. This is not surprising based on what we know about the importance of cost constraints on brain organisation \cite{bassett2010efficient,bullmore2012economy,bassett2009cognitive,fornito2011genetic}. Strong connectivity weights indicate a large number of axonal projections, a big bandwidth bundle, perhaps macroscopically visible as a white matter tract. Building and resourcing a high bandwidth axonal signalling bundle is a significant biological cost that will increase as a function of connection distance: it is parsimonious to wire high bandwidth over short distances. Short distance connections are not only strongly weighted but also topologically clustered. So the strongest weights in both cortical networks define a topologically segregated and anatomically localised organisation. A map of the sub-network formed by the strongest weights shows spatial and topological clusters of regions (Fig.~\ref{WeakLinks}). In the mouse, the strongly weighted clusters each comprise functionally specialised areas of cortex (visual, motor, etc.) that are known to be densely inter-connected and anatomically localised \cite{rubinov2015wiring,ypma2016}. Thus it is not surprising that weighting the topological analysis of mammalian cortical networks will provide stronger evidence for non-random clustering  than unweighted analysis of binary graphs.

The most weakly weighted connections are an area of active, ongoing research (discussed in more detail below) and it is inevitable that there is still much to learn about a feature of network organisation -- replicable but very weak connections between large cortical areas -- that had not been measurable until recent advances in tract-tracing methodology. However, it is clear that weaker connections tend to subtend longer distances, and can be either more topologically random than \cite{ypma2016} or similarly topologically organized to \cite{bassett2012altered} strong connections.

We conclude that graph theoretical analysis of tract-tracing connectomes should respect the quality of the data and use weighted topological metrics to reflect the wide ranging variation in anatomical connectivity, from single fibres to major tracts, that is now measurable in the mammalian brain \cite{wang2016brain}. Weighted graph analysis demonstrates clearly that both the macaque and mouse connectomes are small-world networks, as are the human, cat, and nematode \cite{muldoon2016small}.  Binary graph analysis has usefully measured high connection density, due to the existence of many new anatomical connections, but binarization of these data is not the best way to understand their complex topology and its economical embedding in anatomical space \cite{rubinov2011weight,bassett2011conserved,klimm2014resolving,bassett2012altered,rubinov2015wiring}. Future studies will likely also pay more attention to the fact that most tract-tracing markers are axonally transported only in one direction: anterograde or retrograde. This means that the weight matrix could be modelled more completely as a weighted and directed graph, representing a further evolution in the use of graph theoretical methods to capture a richer and biologically more meaningful model of brain network organisation than can be provided by binary graphs of unweighted and undirected edges.

\section*{The utility of weak connections}

At this juncture, one might naturally ask: ``From a neuroscientific perspective, do we need techniques that account for edge weights? Do these weights indeed capture information of relevance for cognition and behavior?'' Neuroanatomical data suggest that the weights of structural connections may be driven by developmental growth rules \cite{klimm2014resolving,lohse2014resolving,kaiser2006nonoptimal,ercsey2013predictive,markov2013cortical}, energetic and metabolic constraints \cite{bassett2010efficient}, and physical limitations on the volume of neural systems, particularly brains encapsulated by bone \cite{sherbondy2009think}. Yet the role of these edge weights in neural computations \cite{schneidman2006weak} and higher order cognition has been less well studied.

Recent studies have begun to elucidate the role of edge weights -- and particularly of weak connections -- in human cognition. In resting state fMRI data, weak functional connections from lateral prefrontal cortex to regions within and outside the frontoparietal network have been shown to display individual differences in strength that predict individual differences in fluid intelligence \cite{cole2012global}. The same general relationship was observed in a separate study in which individual differences in moderately weak, long-distance functional connections at rest were strongly correlated with full scale, verbal, and performance IQ \cite{santarnecchi2014efficiency}. Neither of these correlations were observed when considering strong connections. Indeed the utility of weak edges appears to extend to psychiatric illness, where the highly-organized topology of weak functional connections -- but not strong functional connections -- in resting state fMRI were able to classify people with schizophrenia from healthy controls with high accuracy and specificity \cite{bassett2012altered}. Interestingly, individual differences in these weak connections were significantly correlated with individual differences in cognitive scores and symptomatology. Together these results demonstrate that, indeed, methods that are sensitive to the strength (or weakness) of individual connections are imperative for progress to be made in understanding individual differences in cognitive abilities, and their alteration in psychiatric disease.

Importantly, the utility of weak connections is not only evident at the large scale in human brains, but also at the neuronal scale as measured in non-human species. In an influential paper published in 2006 with Bialek and colleagues, Schneidman demonstrated that weak pairwise correlations implied strongly correlated network states in a neural population, suggesting the presence of strong collective behaviour \cite{schneidman2006weak}. This result was initially counter-intuitive as one might expect that weak correlations would be associated with the lack of collective behavior. However, the original observation has withstood the test of time, and has been validated in several additional studies including work at the level of tract tracing in macaque monkeys \cite{goulas2015strength}. Intuitively, the juxtaposition of weak correlations and cohesive, collective behavior is thought to be driven by the underlying sparsity of neuronal interactions \cite{ganmor2011architecture}, which contain a few non-trivial higher-order interaction terms \cite{ganmor2011sparse}. Indeed, these higher-order interactions are the topic of some interest both from a computational neuroscience perspective \cite{giusti2016twos,sizemore2016classification}, and from the perspective of neural coding \cite{giusti2015clique}.

But perhaps the claim that weak connections are critically important for our understanding of neural systems should not be particularly surprising. Indeed, it is in fact an old story, first  published at the inception of network science. In 1973, Granovetter wrote a seminal paper, titled ``The strength of weak ties'', which highlighted the critical importance of weakly connected components in global system dynamics \cite{granovetter1973strength}. Such weak connections are ubiquitous in many systems, from physician interactions \cite{bridewell2011social} to ecosystem webs \cite{ulanowicz2014limits} and atmospheric pathways \cite{lee2014tracking}.  Looking forward, critical open questions lie in how these weak connections drive global dynamics, and how one can intervene in a system to manipulate those processes \cite{gu2015controllability,betzel2016optimally,muldoon2016stimulation}.

Acknowledging the role of weak connections, weighted small-world organization plays a critical role in system functions that are particularly relevant to neural systems: including coherence, computation, and control and robustness \cite{novkovic2016topological}. Perhaps the most commonly studied function afforded by small-world architecture is the ability to transmit information, a characteristic that is common in networks of coupled oscillators \cite{barahona2002synchronization,hong2002synchronization,nishikawa2003heterogeneity} (although see \cite{atay2006synchronization} for a few notable exceptions). This capability supports enhanced computational power \cite{lago2000fast}, via swift flow and transport \cite{hwang2010spectral}. In dynamic networks, oscillators coupled on small-world networks are much more sensitive to link changes than their random network counterparts \cite{kohar2014synchronization}, the time taken to reach synchronization is lowered, and the synchronized state is less stable over time, potentially enabling greater diversity of function. When such a system has both small-world topology and geometry, it directly impacts the network's ability to speed or slow spreading \cite{karsai2011small}, a potentially useful characteristic for resilience to dementia which is thought to be caused by the spread of prions \cite{raj2012network,raj2015network}.

The value of small-world architecture is not limited to its support of synchronization and information flow. Instead, it also supports a wide-range of computations in neural circuits. From early neural network studies, it is clear that the exact topology of connectivity patterns between network elements directly supports tradeoffs in the network's ability to learn new information \emph{versus} retain old information in memory  \cite{hermundstad2011learning}. When these patterns are organized in a small-world manner, evidence suggests that local computations can be integrated across distributed cell assemblies to support functions as diverse as somatosensation \cite{zippo2013neuronal} and olfaction \cite{imam2012implementation}. The mechanism by which small-worlds support these computations may stem from the fact that their topological structure tends to contain both large cavities and high-dimensional cliques \cite{sizemore2016classification}, which when embedded in a physical space can strongly constrain the geometric properties of the computation \cite{giusti2015clique}.

While small-world structure can offer non-trivial advantages in terms of both communication and computation, it also directly informs the sorts of interventions that one could use to guide network dynamics and by extension system function. Indeed, computational studies have demonstrated that small-world network architecture requires specific control strategies if one wishes to stem the propagation of seizure activity \cite{ching2012distributed}, control the spread of viruses \cite{kleczkowski2012searching}, or enhance recovery following injury \cite{hubler2008mathematical}. To gain an intuition for how topology impacts control, we can consider the broad-scale degree distribution also characteristic of brain networks. Based on the Laplacian spectrum, one can observe that weakly connected nodes have the greatest potential to push the system into distant states, far away on an energy landscape \cite{pasqualetti2014controllability}; conversely, strongly connected hubs have the greatest potential to push the system into many local states, nearby on the energy landscape \cite{gu2015controllability}. Thus, control energy (such as that provided by brain stimulation) may be targeted to different locations in a small-world brain network to affect a specific change in brain dynamics \cite{muldoon2016stimulation}. 

\section*{Conclusions}
Small-worldness remains an important and viable concept in network neuroscience. Nearly twenty years on from the first analysis of the complex topology of a binary graph representing the nervous system of {\em C. elegans}, it has been established that small-worldness is a nearly-universal and functionally valuable property of nervous systems economically embedded in anatomical space. Recent advances in tract tracing connectomics do not refute small-worldness; rather they considerably enrich and deepen our understanding of what it means in the brain. The extraordinary precision of contemporary tract tracing, and the important discovery that mammalian cortical networks are denser than expected, mandates the adoption of more sophisticated techniques for weighted graph theoretical modelling of inter-areal connectomes. On this basis, we expect the next ten years to yield further insights into the functional value of weak as well as strong connections in brain networks with weighted small-worldness.

\section*{Acknowledgements and disclosures}

We thank Rolf Ypma and Evelyn Tang for comments on an earlier draft of this manuscript, and for Jonathan Soffer for assistance. ETB is employed half-time by the University of Cambridge and half-time by the University of Cambridge; he holds stock in GSK. DSB acknowledges support  from the John D. and Catherine T. MacArthur Foundation, the Alfred P. Sloan Foundation, the  Army Research Laboratory and the Army Research Office through contract numbers W911NF-10-2-0022 and W911NF-14-1-0679,the National Institute of Mental Health (2-R01-DC-009209-11), the National Institute of Child Health and Human Development  (1R01HD086888-01), the Office of Naval Research, and the National Science Foundation (BCS-1441502, BCS-1430087, PHY-1554488, and BCS-1631550).

\bibliographystyle{naturemag}
\bibliography{SWR}

\end{document}